\documentclass{aa}  
\usepackage{graphicx}
\usepackage{balance}
\usepackage{txfonts}
\begin{document}

   \title{The extended molecular envelope of the asymptotic giant branch star $\pi^{1}$ Gruis as seen by ALMA}

   \subtitle{I. Large-scale kinematic structure and CO excitation properties}

   \author{L. Doan
          \inst{1}
          \and
          S. Ramstedt\inst{1} \and W.~H.~T. Vlemmings\inst{2} \and S. H\"ofner\inst{1} 
          \and 
          E. De Beck\inst{2} \and F. Kerschbaum\inst{3} \and M. Lindqvist\inst{2} \and M. Maercker\inst{2} \and S. Mohamed\inst{4,5,6} \and C. Paladini\inst{7} \and M. Wittkowski\inst{8}
          }

   \institute{Department of Physics and Astronomy, Uppsala University, Box 516, 751 20 Uppsala, Sweden\\
              \email{lam.doan@physics.uu.se} \and Department of Earth and Space Sciences, Chalmers University of Technology, SE-43992 Onsala, Sweden \and Department of Astrophysics, University of Vienna, T\"urkenschanzstr. 17, 1180 Vienna, Austria \and South African Astronomical Observatory, P.O. Box 9, 7935 Observatory, South Africa \and Astronomy Department, University of Cape Town, University of Cape Town, 7701, Rondebosch, South Africa \and  National Institute for Theoretical Physics, Private Bag X1, Matieland, 7602, South Africa \and Institut d'Astronomie et d'Astrophysique, Universit\'e Libre de Bruxelles, Campus Plaine C.P. 226, Boulevard du Triomphe, B-1050 Bruxelles, Belgium \and European Southern Observatory, Karl-Schwarzschild-Stra\ss e 2, 85748 Garching, Germany}

   \date{}

% \abstract{}{}{}{}{} 
% 5 {} token are mandatory
 
  \abstract
   {The S-type asymptotic giant branch (AGB) star $\pi^{1}$~Gru has a known companion at a separation of 2\farcs7 ($\approx$400 AU). Previous observations of the circumstellar envelope (CSE) show strong deviations from spherical symmetry. The envelope structure, including an equatorial torus and a fast bipolar outflow, is rarely seen in the AGB phase and is particularly unexpected in such a wide binary system. Therefore a second, closer companion has been suggested, but the evidence is not conclusive.}
   {The aim is to make a 3D model of the CSE and to constrain the density and temperature distribution using new spatially resolved observations of the CO rotational lines.} 
   {We have observed the $J$=3-2 line emission from $^{12}$CO and $^{13}$CO using the compact arrays of the Atacama Large Millimeter/submillimeter Array (ALMA). The new ALMA data, together with previously published $^{12}$CO $J$=2-1 data from the Submillimeter Array (SMA), and the $^{12}$CO $J$=5-4 and $J$=9-8 lines observed with Herschel/Heterodyne Instrument for the Far-Infrared (HIFI), is modeled with the 3D non-LTE radiative transfer code SHAPEMOL.}
   {The data analysis clearly confirms the torus-bipolar structure. The 3D model of the CSE that satisfactorily reproduces the data consists of three kinematic components: a radially expanding torus with velocity slowly increasing from 8 to 13\,km\,s$^{-1}$ along the equator plane; a radially expanding component at the center with a  constant velocity of 14\,km\,s$^{-1}$; and a fast, bipolar outflow with velocity proportionally increasing from 14\,km\,s$^{-1}$ at the base up to 100\,km\,s$^{-1}$ at the 
tip, following a linear radial dependence. The results are used to estimate an average mass-loss rate during the creation of the torus of 7.7$\times$10$^{-7}$\,M$_{\odot}$\,yr$^{-1}$. The total mass and linear momentum of the fast outflow are estimated at 7.3$\times$10$^{-4}$\,M$_{\odot}$ and 9.6$\times$10$^{37}$\,g\,cm\,s$^{-1}$, respectively. The momentum of the outflow is in excess (by a factor of about 20) of what could be generated by radiation pressure alone, in agreement with recent findings for more evolved sources. The best-fit model also suggests a $^{12}$CO/$^{13}$CO abundance ratio of 50. Possible shaping scenarios for the gas envelope are discussed.}
   {}


   \maketitle
%
%________________________________________________________________

\section{Introduction}
Asymptotic giant branch (AGB) stars are believed to evolve from low- to intermediate-mass (0.8--8\,M$_{\odot}$) main sequence stars. The evolution of AGB stars  is governed by the massive wind from the stellar surface \citep[typical expansion velocity, $v_{\rm{exp}}$ $\sim$ 10\,km\,s$^{-1}$, and mass-loss rate, $\dot{M}$ $ \backsim $ 5$\times$10$^{-7}$\,M$_{\odot}$\,yr$^{-1}$; e.g.,][]{ramsetal09}, which creates an expanding circumstellar envelope (CSE) of molecular gas and dust already on the AGB. The physical processes behind the AGB-star wind are comparatively well understood and current radiation-hydrodynamical models reproduce observed properties well \citep{eriketal14,bladetal15}. However,  several important aspects still need to be investigated further, such as the wind evolution over time and the formation of complex large-scale structures, to establish the formation scenario
for planetary nebulae (PNe).

Planetary nebulae are large, tenuous emission nebulae glowing at visible wavelengths through recombination and forbidden lines from ionized (atomic) gas. Expansion velocities of PNe are on average about a factor of two larger \citep[$\sim$20\,km\,s$^{-1}$;][]{gesizijl00,huggetal05} than typical AGB wind velocities, but more extreme velocities in excess of $\sim$100\,km\,s$^{-1}$ are also found in post-AGB stars and PNe \citep[e.g.,][]{vlemetal06,clynetal15}. \citet{bujaetal01} studied the CO emission from 30 protoplanetary nebulae (P-PNe) and found that almost all of the sample sources have both a slowly expanding envelope (probably the remnant AGB wind) and fast (often bipolar) outflows. The momenta of the envelopes are consistent with a wind driven by radiation pressure on dust grains (as on the AGB), but an alternative mechanism is necessary to accelerate fast outflows. Imaging surveys of PNe \citep[see, e.g.,][and references therein]{saha14} show that less than 5\% of PNe are round, as would be expected if they were the direct result of an isotropic wind, while a majority of the AGB envelopes seem to be spherically symmetric \citep{castetal10} on large scales. This further supports the idea that strong dynamical evolution takes place before AGB stars become PNe. While different shaping agents (binary interaction, large planets, and global magnetic fields) have been suggested \citep[see][]{balifran02}, much work remains before the theoretical models can be confirmed observationally. 

In this context, molecular line observations of transitional objects are extremely valuable since they trace the remnant material of the AGB wind and give kinematic information \citep[e.g.,][]{sancsaha12}. The unsurpassed potential of the Atacama Large Millimeter/submillimeter Array (ALMA) to study these transitional objects has already been demonstrated with Early Science capabilities \citep{bujaetal13,olofetal15}. To add to previous studies, we observed four binary stars on the AGB (R~Aqr, Mira, W~Aql, and $\pi^{1}$~Gru) with ALMA to investigate the dependence of the circumstellar shaping and morphology on the AGB on the binary separation and wind properties. The observations of Mira ($o$~Cet) already revealed how the massive AGB wind has been sculpted by the fast, thinner wind from the companion \citep{ramsetal14}, and in this paper, we present the initial results on the largest separation source of our sample, $\pi^{1}$~Gru.  

 $\pi^{1}$~Gru is an evolved, S-type AGB star \citep{vecketal98} at a distance of about 150\,pc \citep{per97} and log(L/L$_{\odot}$)=3.86 \citep{vanEetal98}. A fast bipolar molecular outflow was discovered by \citet{saha92} in $^{12}$CO $J$=1-0 and 2-1 emission. \citet{saha92} also observed the $^{13}$CO $J$=1-0 line emission and found a $^{12}$C/$^{13}$C abundance ratio in the range 25-50. $\pi^{1}$ Gru has a known G0V companion \citep{feas53,akejohn92} at 2\farcs7 separation, but  \citet{saha92} already mentioned that a closer unknown companion would be required to explain the observed morphology. The $^{12}$CO $J$=2-1 emission was mapped by \citet{knapetal99} and later by \citet{chiuetal06} using the SMA (with a synthesized beam of $2\farcs2\times4\farcs2$). \citet{chiuetal06} built on the model by \citet{knapetal99} and suggested that the star is surrounded by a thick, low-velocity (11\,km\,s$^{-1}$) expanding torus with a faster bipolar outflow that is oriented perpendicular to the torus; this torus is referred to as a flared disk, but we use torus throughout
this paper, since the structure shows no sign of rotating. Herschel/PACS observations show a large arc, possibly a spiral arm, reaching out at $\sim$40\arcsec to the east of the star \citep{mayeetal14} and possibly shaped by the known 2\farcs7 companion. \citet{mayeetal14} have also analyzed VLTI/MIDI and  AMBER data, together with Hipparcos and Tycho observations to search for a closer companion. Although the second companion is not directly detected in the VLTI observations, they have found support for a closer companion (at 10-30\,AU separation) from the combined analysis of the available data. 

We observed  $\pi^{1}$~Gru in $^{12}$CO and $^{13}$CO $J$=3-2 with ALMA. In this paper we construct a 3D kinematic and radiative transfer model based on the ALMA Atacama Compact Array (ACA) and Total Power (TP) observations, together with the previously published $^{12}$CO $J$=2-1 SMA observations from \citet{chiuetal06} via the publicly available radiative transfer code SHAPEMOL \citep{santetal15}. We present the new ALMA observations and the previously published observations (SMA and Herschel/HIFI) in Section \ref{sec:obs} and the ALMA and SMA observational results in Section \ref{sec:Observationalresults}. The radiative transfer modeling is described in Section \ref{sec:model} and the model results are given in Section \ref{sec:ModelResults}. We give the Discussion and Summary in Sect. 6 and 7.

\section{Observations}
\label{sec:obs}

\subsection{New CO radio line observations with ALMA} 
\label{sec:alma} 
The  $^{12}$CO and $^{13}$CO $J$=3-2 emission was observed with the ALMA-ACA in 2013. The observations consist of a four-point mosaic. They were performed using four spectral windows with a width of 2 GHz each, centered on 331, 333, 343, and 345 GHz. The $u-v$ coverage of an interferometer is always incomplete. Insufficient $u-v$ coverage can cause artificial features when imaging. Observations with the ALMA-TP array was performed in cycle 2 (in 2015) to recover the most extended emission from the source and produce high fidelity images. This array included three 12m antennas, and the observations were performed in single-dish, on-the-fly mapping mode.

\begin{table*}[th]
\centering
\caption{Summary of the interferometric observations and final image cubes.}
\label{tab:observation}
\begin{tabular}{*{8}{c}}
\hline \hline
 Transition  &  Frequency   & Array  & On-source time  & TP On-source time &  $ \Delta \upsilon$  & HPBW & rms \\
                  &      $ \left[ \rm{GHz} \right]  $              &                      & $ \left[ \rm{min} \right]  $                    &  $ \left[ \rm{min} \right]  $                 & $ \left[ \rm{km\,s}^{-1} \right]  $   &               & $ \left[ \rm{Jy\,beam}^{-1} \right]  $ \\
  \hline 
  $^{12}$CO $J$=3-2  &          345.796             &       ALMA-ACA               & 23.7                   &  67.1                & 2     &      $4\farcs56\times2\farcs56$         &   0.051\\                  
 $^{13}$CO $J$=3-2  &           330.588             &       ALMA-ACA               & 23.7                   &  67.1                & 3     &      $5\farcs19\times2\farcs57$         &   0.045\\                  
$^{12}$CO $J$=2-1  &            230.538             &       SMA              & 180                   &  0                & 2     &      $4\farcs20\times2\farcs20$         &   0.114\\                  
 \hline
\end{tabular}
\tablefoot{The spectral resolution $ \Delta \upsilon$ and rms of emission-free channels are from the final image cubes of the combined data. The $^{12}$CO $J$=2-1  image cube does not include TP observations.}
\end{table*}

We used the Common Astronomy Software Application (CASA) for calibration and imaging \citep{casa}. Firstly, the interferometric data was calibrated and preliminary imaged to combine with the TP images later on. Quasars J0006-0623 and J2235-4835 were used as bandpass and complex gain calibrators, respectively. Uranus was used for flux calibration. Because of the low signal-to-noise ratio of the $^{13}$CO $J$=3-2 emission relative to $^{12}$CO $J$=3-2, the $^{13}$CO $J$=3-2 visibility data was imaged using natural weighting to improve the sensitivity. The spectral resolution was about 0.5\,km\,s$^{-1}$, but has been binned to 2\,km\,s$^{-1}$ for $^{12}$CO $J$=3-2 and 3\,km\,s$^{-1}$ for $^{13}$CO $J$=3-2 to improve the signal-to-noise ratio in the images cubes. The resulting line profiles with a recovered flux in the ALMA-ACA observation are discussed in Section \ref{sec:lineprofile}. 

Secondly, the TP observation calibration were carried out with quasars J2230-4416 for focusing, J2230-4416 for pointing, and both of these and $\pi^{1}$~Gru for atmospheric calibration. The brightness of the TP observation was first given in main-beam brightness temperature scale ($T_{\rm{mb}}$) and then converted to Jy\,beam$^{-1}$. The conversion factors of the data from K to Jy\,beam$^{-1}$ are 43.2 at 345 GHz and 45.0 at 330 GHz. The beam size of the TP observation is $19\arcsec$ at 345 GHz and the rms noise level of the images is 0.6\,Jy\,beam$^{-1}$ for $^{12}$CO $J$=3-2 and 0.58\,Jy\,beam$^{-1}$ for $^{13}$CO $J$=3-2. The overall uncertainty of the TP calibration is about 5\%. 

The data was finally combined using CASA packages. Owing to the difference in spatial pixel sizes and spectral ranges between the TP map and the ALMA-ACA map, the TP map was first regridded to the coordinate system of the ALMA-ACA map. Then the ALMA-ACA data was re-imaged using the CLEAN package in an iterative procedure with a decreasing threshold parameter. In this procedure the ALMA-TP map was used as a model for initial cleaning. Finally, the ALMA-TP and ALMA-ACA images were combined via the FEATHER package. A summary of the observations and the final image cubes is given in Table \ref{tab:observation}.

\begin{figure*}[t]
\centering
\includegraphics[width=\hsize]{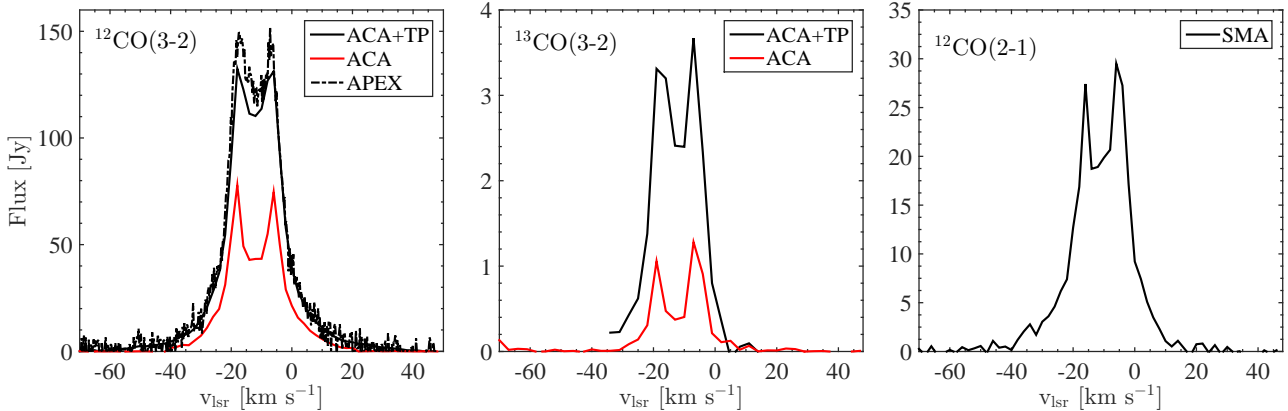}
\caption{Total flux of the $^{12}$CO $J$=3-2 (\textit{left}) and $^{13}$CO $J$=3-2 (\textit{middle}) emission from the ALMA-ACA data (red) and the combined (ACA+TP) data (black). The spectra were generated by convolving the images with the APEX beam (FWHM = $18\arcsec$). The $^{12}$CO $J$=3-2 spectrum \citep{rams06} from APEX (FWHM = $18\arcsec$) is plotted for comparison. Owing to the artificial feature (explained in Sect. \ref{sec:lineprofile}) in the blueshifted wing of the $^{13}$CO $J$=3-2 line, the line profile of the combined data was only plotted from -35\,km\,s$^{-1}$. The $^{12}$CO $J$=2-1 spectrum (\textit{right}) from the SMA observation \citep{chiuetal06} was generated by convolving the image with the SMA primary beam of $55\arcsec$. }
\label{fig:TP_spectrum}
\end{figure*}

\subsection{Previously published CO radio line observations} 
\label{sec:sma} 

\textbf{SMA Observations:} The $^{12}$CO $J$=2-1 observation was performed in 2004 using the SMA with a 2 GHz bandwidth correlator and a 812.5 kHz channel separation over 256 channels. The source was observed with 28 baselines and the longest baseline was about 83$\,\rm{k\lambda}$. The calibrators were observed along with the target. The data has previously been published by \citet{chiuetal06}. 

In this study, the upper sideband data has been recalibrated and reimaged using CASA. The bandpass calibration was carried out using Uranus. The nearby quasar 2258-279 was set as the complex gain calibrator and absolute flux calibrator (instead of Uranus as in \cite{chiuetal06}). We applied the Briggs weighting method for imaging and an active mode for cleaning the dirty maps in which the emitting region was carefully selected to avoid the strong noise speaks. This resulted in some differences in the line profile compared to that found in \cite{chiuetal06} (see Section \ref{sec:lineprofile}). The observations and the final image cube are summarized in Table \ref{tab:observation}.

\textbf{Herschel/HIFI observations:} The $^{12}$CO $J$=5-4, 9-8, and 14-13 observations were part of the Herschel SUCCESS program \citep{teys11,dani15} and observed with the onboard instrument HIFI \citep{graa10}. The signal-to-noise ratio of the $^{12}$CO $J$=14-13 line was about 2 to 3 and because of this high uncertainty we decided to omit this line from our analysis. The telescope
beam sizes of $^{12}$CO $J$=5-4 and 9-8 observations are $36\farcs1$ and $20\farcs1$, respectively. The technical setup, data reduction, and line profiles were published in \cite{dani15}. The line profiles were plotted in main beam temperature scale with a noise rms of 15 mK at a channel resolution of 3\,km\,s$^{-1}$. In this paper, we adopted the line profiles without any recalibration.   

\section{Observational results and discussion}
\label{sec:Observationalresults} 
\subsection{The continuum flux}
\label{sec:continuum}

A single continuum source was detected from the line-free channels of the ALMA-ACA and SMA observations. Neither the known companion, nor  a closer companion, such as that proposed by Mayer et al. (2014), would be resolved. The continuum flux densities are 32.7$ \pm $6 mJy at 230 GHz and 82.2$ \pm $4.7 mJy at 343 GHz. Assuming  optically thick blackbody emission from a stellar photosphere with a temperature of 3000 K \citep{vant02,abia98} and a stellar radius of 2.2$\times$10$^{13}$ cm \citep{chiuetal06}, the flux densities would be 35 mJy and 78 mJy, respectively. The measured continuum flux is consistent with the thermal emission of the stellar photosphere. Since the stellar temperature, radius, and flux measurement are highly uncertain, it is not enough to determine whether there is a contribution from dust continuum emission.

\subsection{Line profiles}
\label{sec:lineprofile}
Fig. \ref{fig:TP_spectrum} (\textit{left and middle}) shows the line profiles of the combined (ACA+TP) $^{12}$CO $J$=3-2 and $^{13}$CO $J$=3-2 data generated by integrating over a circular region with the width of the APEX beam (18$\arcsec$) centered on the stellar position. The $^{12}$CO $J$=3-2 line profile from APEX \citep{rams06} is also plotted to evaluate the recovered flux in the combined ALMA data. A comparison shows that the ALMA-ACA observations recovered a fraction of less than 50\% of the flux observed in the TP observation. The combined ALMA maps contain $ \backsim $90\% of the flux observed by APEX in 2005, which is within the calibration uncertainties. The lower sideband containing the $^{13}$CO line was affected by the mirrored $^{12}$CO line from the upper sideband. This resulted in an artificial feature in the blueshifted wing of the $^{13}$CO $J$=3-2 line profile. The line is therefore only plotted from -35\,km\,s$^{-1}$ in Fig.~\ref{fig:TP_spectrum} (\textit{middle}). The $^{12}$CO $J$=2-1 emission observed with the SMA \citep{chiuetal06}, shown in Fig. \ref{fig:TP_spectrum} (\textit{right}), is convolved with the SMA primary beam (FWHM of $55\arcsec$). 

\begin{figure*}
\centering
\includegraphics[width=\hsize]{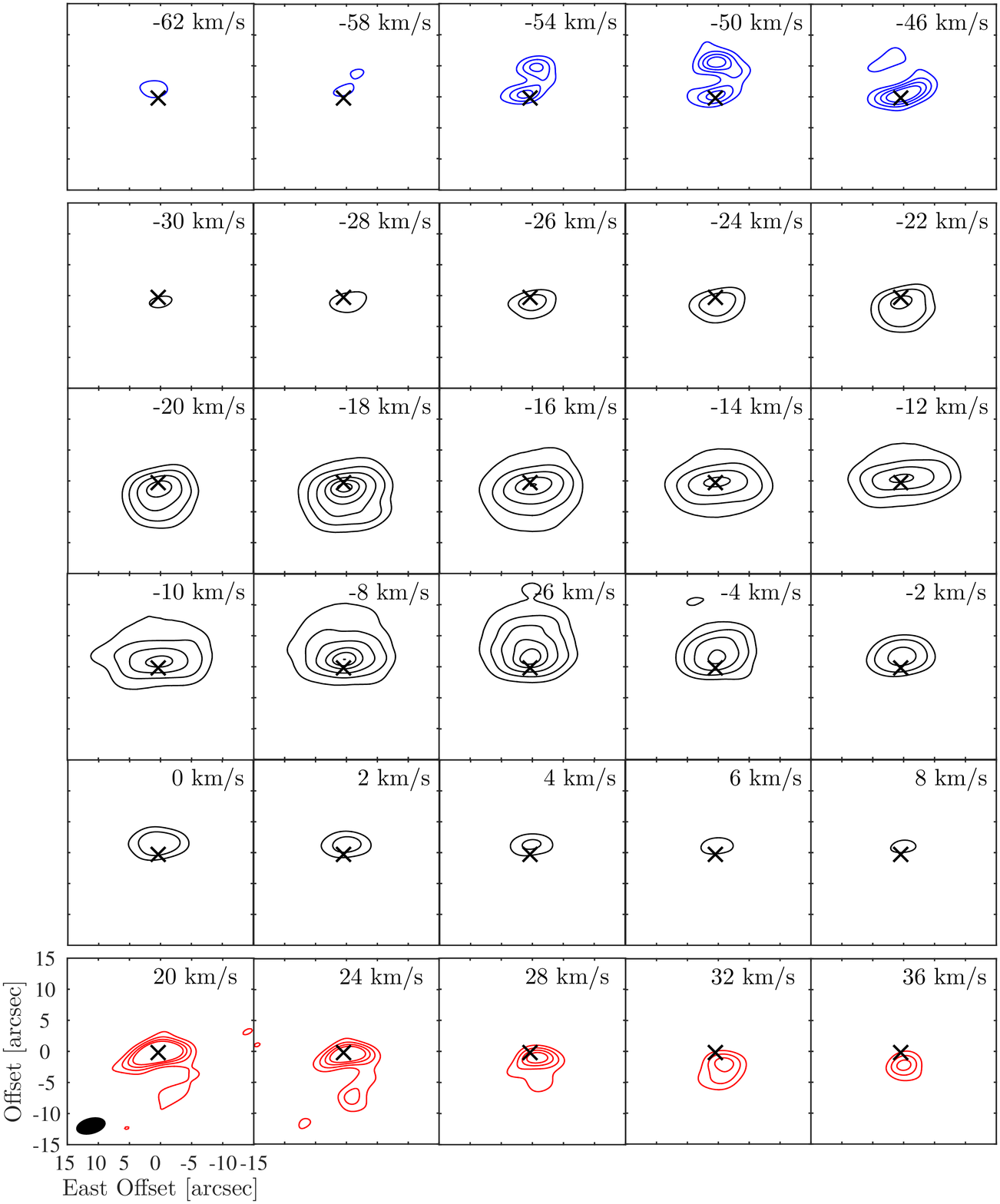}
\caption{Contour maps of  the $^{12}$CO $J$=3-2 emission from the combined ALMA data. Contour levels are at 3, 5, 9, 15, 20, 25, and 30$\sigma$ ($\sigma$=1.2\,Jy\,beam$^{-1}$ for the middle rows, derived from the plotted channels), and at 3, 5, 7, and 9$\sigma$ ($\sigma$=0.08\,Jy\,beam$^{-1}$ for the first and last row, derived from the plotted channels). The synthesized beam size of $4\farcs56\times2\farcs56$ at a PA of -81$^{\circ}$ is plotted in the lower left corner of the 20\,km\,s$^{-1}$ channel. The local standard of rest velocity is given in the upper right corner of each channel. A cross denotes the stellar position determined from the continuum emission.}
\label{fig:maps12CO32}
\end{figure*}

\begin{figure*}
\centering
\includegraphics[width=\hsize]{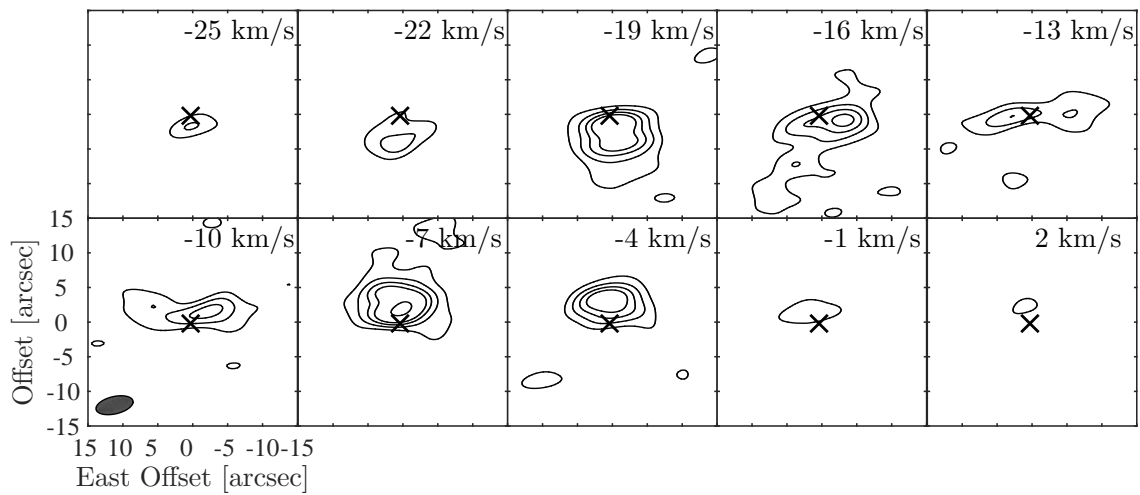}
\caption{Same as Fig. \ref{fig:maps12CO32}, for the $^{13}$CO $J$=3-2 emission from the combined ALMA data. Contour levels are at 3, 5, 7, 9 15, and 20$\sigma$ ($\sigma$=0.055\,Jy\,beam$^{-1}$, derived from the full data set). The synthesized beam size of $5\farcs19\times2\farcs57$ at a PA of -81$^{\circ}$ is plotted in the lower left corner of the -10\,km\,s$^{-1}$ channel.}
\label{fig:maps13CO32}
\end{figure*}

\begin{figure*}
\centering
\includegraphics[width=0.9\hsize]{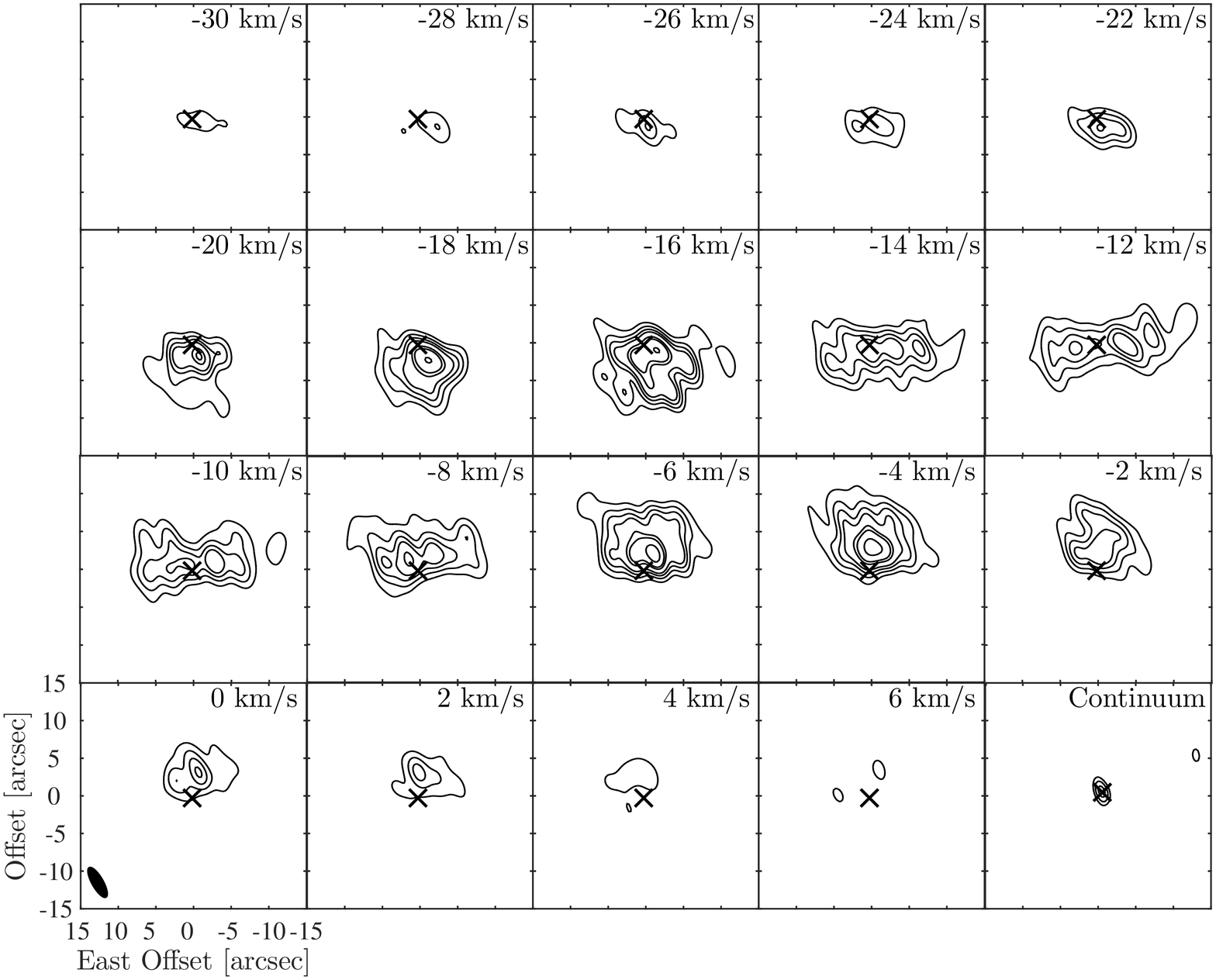}
\caption{Same as Fig. \ref{fig:maps12CO32}, for the recalibrated and reimaged $^{12}$CO $J$=2-1 emission from the SMA \citep{chiuetal06}. Contour levels are at 3, 5, 7, 9, 13, 15, and 20$\sigma$ ($\sigma$=0.3\,Jy\,beam$^{-1}$, derived from the full data set). The synthesized beam is shown in the lower left corner of the 0\,km\,s$^{-1}$ channel and is $4\farcs2\times2\farcs2$ at a PA=-22$^{\circ}$.}
\label{fig:maps12CO21}
\end{figure*}

All emission lines (except the $^{13}$CO $J$=3-2) show a double-horned profile with steep sides at intermediate velocity and extended wings. The double-horned core of the spectral line shows a slowly expanding component, while the high-velocity wings, reaching out to $ \pm $60\,km\,s$^{-1}$ relative to the systemic velocity, are indicative of an outflow that is much faster than a typical AGB wind. There is no sign of an asymmetry in the $^{12}$CO $J$=3-2 data observed with the mosaic field $ \backsim $40$\arcsec$. \cite{chiuetal06} suggested that an over-resolved inhomogeneous structure caused the stronger peak at the blueshifted velocities in the $^{12}$CO $J$=2-1 line profile from their analysis of the data. In contrast, the peak is seen on the redshifted side of both the line profile observed by \cite{knapetal99} and the line profile from our recalibration of the data. This discrepancy may be due to the different calibration and data reduction strategy, as already mentioned in Section \ref{sec:sma}.  

\subsection{Images}
\label{sec:images}

\textbf{$^{12}$CO $J$=3-2:} The channel maps of the $^{12}$CO $J$=3-2 emission from the combined data, shown in Fig. \ref{fig:maps12CO32}, were constructed by integrating over 2\,km\,s$^{-1}$ close to the systemic velocity (from -30\,km\,s$^{-1}$ to 8\,km\,s$^{-1}$) and over 4\,km\,s$^{-1}$ for the high velocities (from -62\,km\,s$^{-1}$ to -46\,km\,s$^{-1}$ and from 20\,km\,s$^{-1}$ to 36\,km\,s$^{-1}$) to increase the signal to noise of the high-velocity channels in which the emission is weaker. 

The emission at the systemic velocity, $V_{\rm{S}} $=-12\,km\,s$^{-1}$, shows what looks like an elongated torus along the east-west (EW) direction. The emission has a maximum close to the stellar position. When moving away from the systemic velocity, the size of the emitting region decreases. Also, the emission gradually moves from the south to the north from blue- to redshifted velocities. This spatial shift of the emission distribution at low relative velocities can be interpreted that the radially expanding torus is inclined relative to the line of sight. 

The high-velocity channel maps are shown in the first and last rows of Fig. \ref{fig:maps12CO32}. In agreement with the $^{12}$CO $J$=2-1 emission \citep{chiuetal06}, the north-south (NS) orientation of the higher velocity emission is opposite to that at lower velocities. Furthermore, the $^{12}$CO $J$=3-2 emission seen in the channels from -58 to -46\,km\,s$^{-1}$ and 20 to 32\,km\,s$^{-1}$ shows an extended region with two separated parts. We propose that this bimodal distribution can be possibly interpreted as emission coming from the lobe walls of a bipolar outflow, while the -62\,km\,s$^{-1}$ and 36\,km\,s$^{-1}$ channels may show the lobe tips or lobe edges at the highest line-of-sight velocity of the bipolar outflow.

\begin{figure*}
\centering
\includegraphics[width=0.9\hsize]{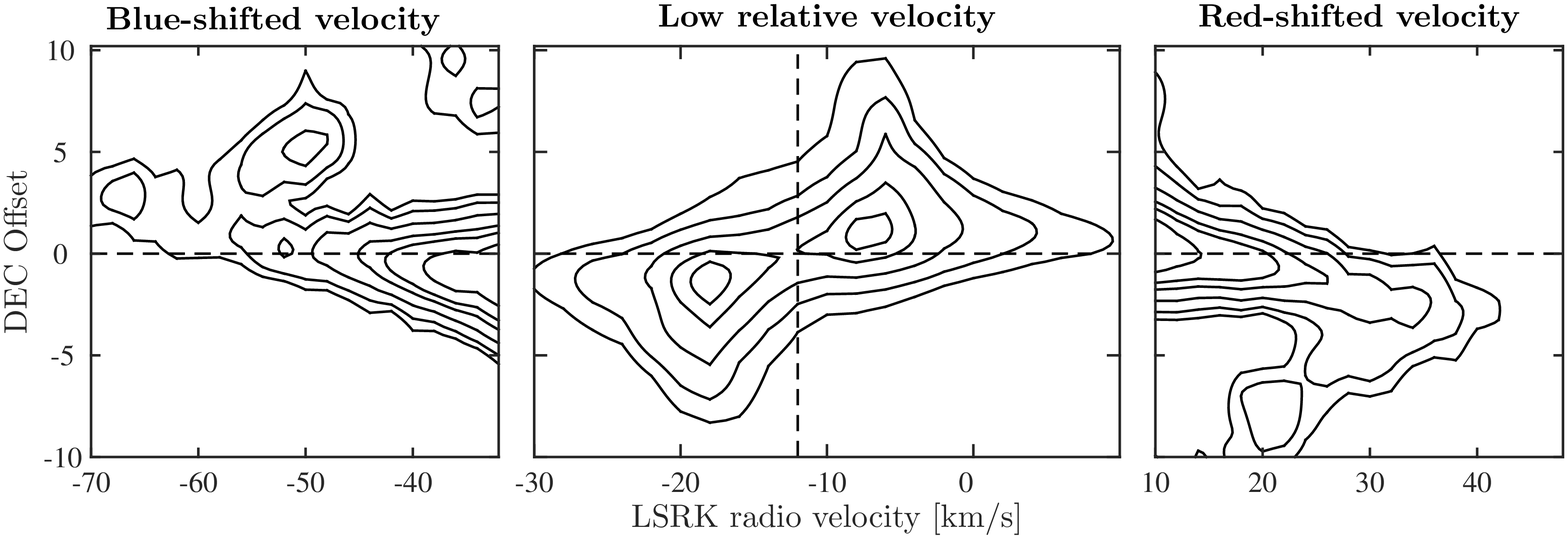}
\caption{The position-velocity diagram for the $^{12}$CO $J$=3-2 emission along a cut at a PA of 0$^{\circ}$ made by integrating over a 2\,km\,s$^{-1}$ velocity interval: low velocity and high signal-to-noise ratio at the middle panels with contour levels at 3, 5, 9, 15, 20, 30$\sigma$ ($\sigma$=1.2\,Jy\,beam$^{-1}$); high velocity and low signal-to-noise ratio at the left and right panel with contour levels at 3, 5, 9, 15, 20$\sigma$ ($\sigma$=0.06\,Jy\,beam$^{-1}$). The vertical and horizontal dashed lines show the systemic velocity ($-$12\,km\,s$^{-1}$) and the stellar position, respectively.}
\label{fig:pv_12CO32}
\end{figure*}

In Fig. \ref{fig:maps12CO32}, as well as in the position-velocity (PV) diagram in Fig. \ref{fig:pv_12CO32}, the emission is divided into two components around -46\,km\,s$^{-1}$ and +20\,km\,s$^{-1}$. At blueshifted velocities, the southern component is just slightly north of the stellar position while the second, northern component is found approximately 5\arcsec  to the north. Assuming that the velocity of the high-velocity outflow increases radially and the system is inclined relative to the line of sight, the gas moving along the edges of lobes have different line-of-sight velocities at the same distance from the equator. For example, the gas moving along the edge of the northern lobe facing the observer, has a higher line-of-sight velocity than the gas moving along the edge away from the observer at the same distance from the equator. This can explain the bimodal distribution seen at, for example -50 km/s. At this velocity channel, the southern emission component would come from the closer lobe edge and the northern emission component would come from the more distant lobe edge, where the same line-of-sight velocity is reached further away from the equator of the system. The exact distribution of the emission as a function of velocity is an intricate function of the gas distribution, i.e., inclination and curvature of the high-velocity outflow and the clumpiness of the gas, and kinematics.

\textbf{$^{13}$CO $J$=3-2:} The $J$=3-2 emission from the less abundant $^{13}$CO isotopologue was imaged by integrating over 3\,km\,s$^{-1}$ (Fig. \ref{fig:maps13CO32}). The emission is very weak compared to the $^{12}$CO  emission and the gas is mostly concentrated and excited in the inner parts of the envelope. At  the systemic velocity, the emission has two peaks on either side of the stellar position along the EW direction. Even if the data was convolved with an identical beam as the $^{12}$CO $J$=3-2, the image would still have this distribution. Any emission is below the noise level at velocities beyond -25\,km\,s$^{-1}$ and 2\,km\,s$^{-1}$.

\textbf{$^{12}$CO $J$=2-1:} The channel maps from the rereduced $^{12}$CO $J$=2-1 line emission are given in Fig.~\ref{fig:maps12CO21}. As previously described \citep{chiuetal06}, the emission at the systemic velocity, $V_{\rm{S}} $=-12\,km\,s$^{-1}$, shows a torus structure that is flared and elongated along the EW direction, similar to the $^{12}$CO J=3-2 distribution. The slightly larger spatial size at each channel velocity compared to the $J$=3-2 emission is caused by the lower minimum kinetic temperature required for the $J$=2-1 excitation. Moreover, the emission has a two-peak (on either side of the stellar position) distribution that differs from the new $^{12}$CO $J$=3-2 data. The gap between the two peaks has been attributed to a central cavity \citep{chiuetal06}, but that is not required to explain the new $^{12}$CO $J$=3-2 data. The orientation of the synthesized beams (see Fig.~\ref{fig:maps12CO21} and \ref{fig:maps12CO32}) can contribute to the different distribution of the $^{12}$CO $J$=2-1 and $^{12}$CO $J$=3-2, but not fully explain the difference. This is discussed further in the following sections.

\subsection{Position-velocity diagrams}
 \label{sec:pv_diagram}
 
The PV diagrams of the $^{12}$CO $J$=3-2 emission shown in Fig. \ref{fig:pv_12CO32} were made along a NS cut (one pixel wide at a PA of 0$^{\circ}$) through the stellar position. A different correlation between the velocity vector field and the position vector is seen for the equatorial torus and the fast bipolar outflow, separately. 

The NS PV $^{12}$CO $J$=3-2 diagram is similar to that of $^{12}$CO $J$=2-1 \citep[see][]{chiuetal06}, but less extended in spatial offset than the $^{12}$CO $J$=2-1 emission (as already seen in the channel maps). The emission at low relative velocity  (plotted in the central plot of Fig.~\ref{fig:pv_12CO32}) is consistent with an expanding, inclined equatorial torus, extended to the north at redshifted velocity, and to the south at blueshifted velocity. 

The emission at higher velocities originates from the fast outflow component (plotted in the far left and right plot of Fig.~\ref{fig:pv_12CO32}) and shows the opposite pattern: a redshifted velocity to the south and blueshifted velocity to the north. In this part of the figure, the position is directly proportional to the velocity, i.e., emission with higher velocity comes from a position further away from the center. The offset emission regions seen at blue- and redshifted velocities (with a Dec-offset beyond about $\pm$5$\arcsec$) correspond to the bimodal distributions also seen in the channel maps in Fig. \ref{fig:maps12CO32}. We do not detect significant emission beyond -70\,km\,s$^{-1}$ in the blueshifted part and +40\,km\,s$^{-1}$ in the redshifted part. This corresponds to a maximum projected gas velocity of about 60\,km\,s$^{-1}$ in the CSE.  

\section{Circumstellar model}
\label{sec:model} 
\subsection{Geometry and velocity field} 
\label{sec:geo} 
\begin{figure}
\centering
\includegraphics[width=\hsize]{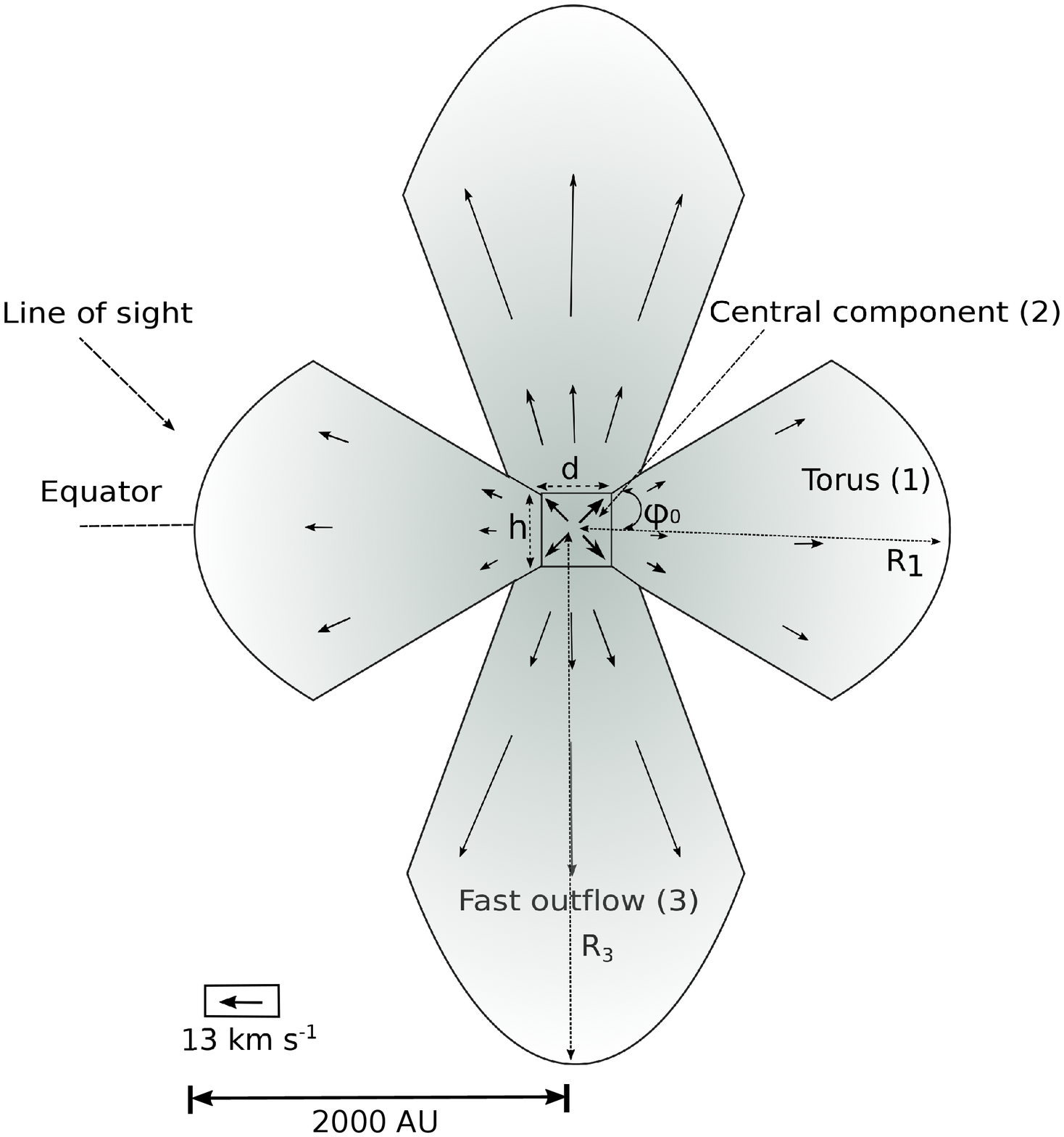}
\caption{Sketch illustrating the three components used to model the CSE of $\pi^{1}$~Gru.}
\label{fig:structure}
\end{figure}

As mentioned above, \citet{chiuetal06} built on the flared-disk model to reproduce the low-velocity component seen in the $^{12}$CO $J$=2-1 SMA maps, but they never attempted to model the fast component. In this study, we reconstructed the gas envelope using both a low-velocity torus and a fast bipolar outflow to study the  full 3D morphology and kinematics of the system. The combined ALMA data are used as observational constraints, together with the previously published SMA data. The structure of the modeled gas envelope is schematically illustrated in Fig. \ref{fig:structure}. The modeled envelope was constructed as a system with the following three components: 
\\(1) A radially expanding torus in the shape of a flared disk with an opening angle of $ 2\varphi_{0} $, an inner radius of $ d/2 $, and an outer radius of $ R_{\rm{1}} $. The radius $ R_{\rm{1}} $ is just the maximum radius used in the model computation. It does not necessarily represent a density-cutoff radius, i.e., the physical outer boundary of the torus; nebular layers beyond $ R_{\rm{1}} $  do not contribute significantly to the observed emission for the density and temperature laws adopted in our model. The outer radius was constrained by the observed angular size at the central channel. The radial gas velocity inside the torus depends on the distance from the center and latitude above or below the equator, 
\begin{equation} \label{eq:eqq1}
\centering
 v_{1}=\left[ v_{\rm{1a}}+v_{\rm{1b}} \frac{r}{R_{1}}\right]f_{\varphi},
\end{equation}
where the constants $  v_{\rm{1a}} $ and $ v_{\rm{1b}} $ were chosen to produce line profiles with the same width as the line cores of the observational data and $ r $ is the radial distance from the center ($ r=\sqrt{x^{2}+y^{2}+z^{2}} $). The $ f_{\varphi} $ factor increases linearly with latitude. It has the value of $ f_{0^{\circ}} =1 $ at the equator and the values of $f_{\pm\varphi_{0}}  $ at the top and bottom edges were determined by fitting the data.  
\\(2) A central, radially expanding component originating at the center and placed inside the torus. The velocity is constant, $ v_{2} $. Its shape is cylindrical with diameter $ d $ and height $ h $. 
\\(3) A faster bipolar outflow perpendicular to the torus extends from the central component. The lobes have a radially expanding velocity field linearly increasing with the distance from the equator,
 \begin{equation} \label{eq:eqq2}
\centering
 v_{3}= v_{\rm{3a}}+v_{\rm{3b}} \frac{z}{R_{3}},
\end{equation}
where  $ v_{\rm{3a}}$ and  $ v_{\rm{3b}}$  were chosen so that the velocity increases from the central component velocity to the highest velocity inferred from the data and $ z$ is the vertical distance from the equator. The radius $ R_{\rm{3}} $ is just the maximum radius used in the model computation. It is not the physical outer boundary of the outflow. The   $ R_{\rm{3}} $ value was poorly constrained by the observations because of the low signal-to-noise  ratio at very high velocities and the dependence on the inclination of the system.  

\begin{figure*}
\centering
\includegraphics[width=\hsize]{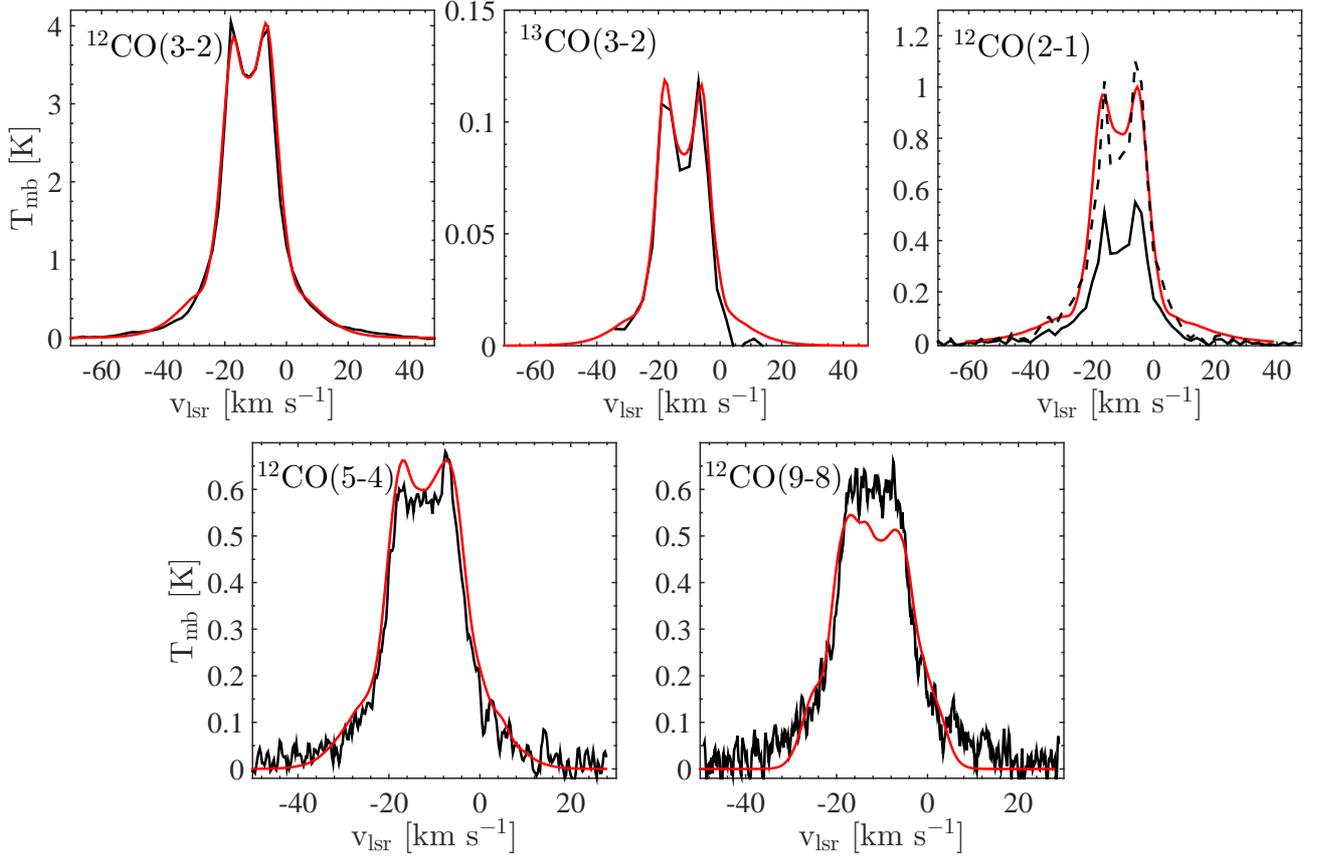}
\caption{Line profiles from the observations (\textit{black}) and the model (\textit{red}). The black dashed line in the upper right panel shows the observed profile multiplied by a factor of 2 to correct for the missing flux according to \cite{chiuetal06}.}
\label{fig:Spectrum_comp}
\end{figure*}

\begin{figure*}
\centering
\includegraphics[width=\hsize]{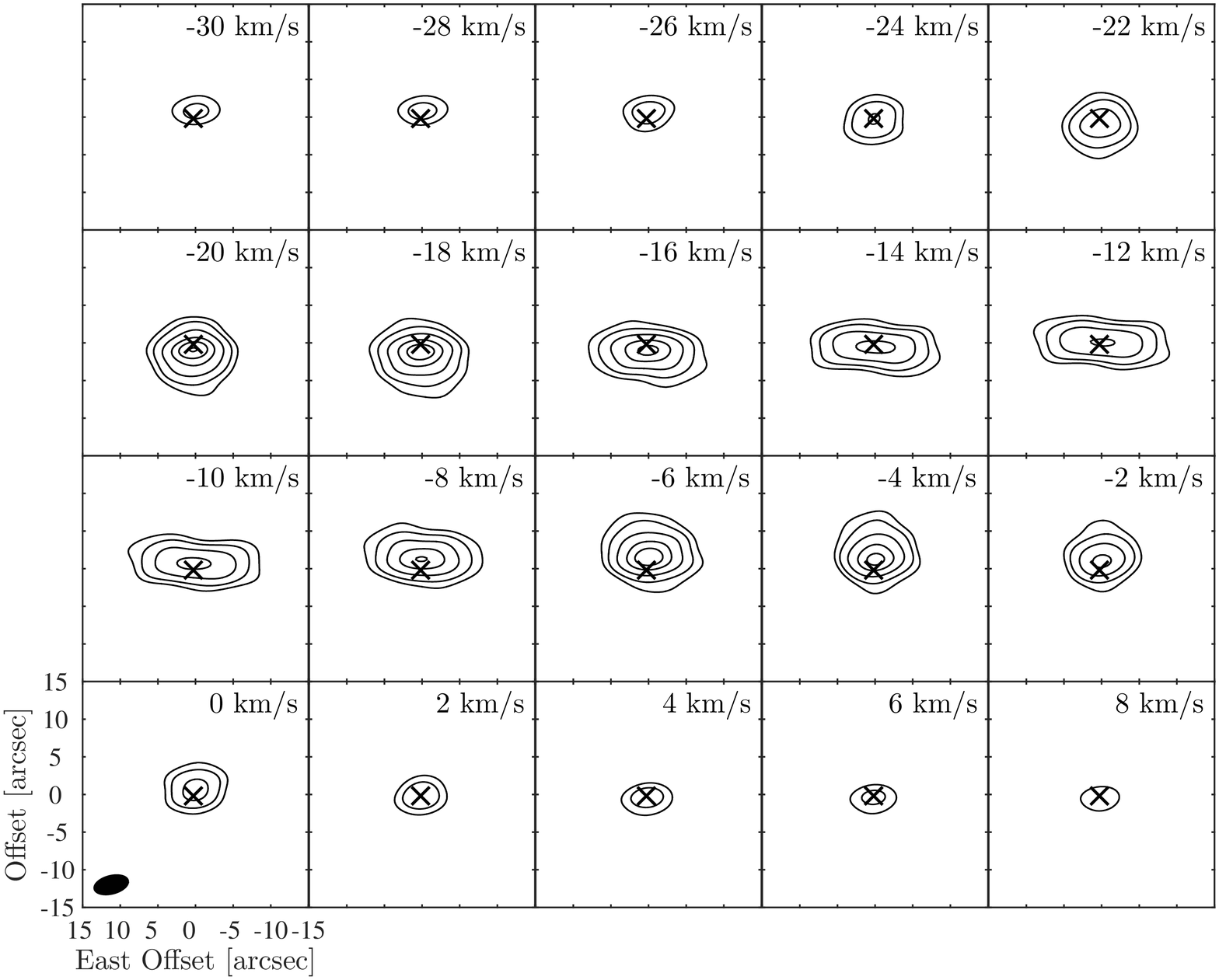}
\caption{Contour maps of the $^{12}$CO $J$=3-2 emission from the model simulating the combined (ACA+TP) ALMA observation. The contour levels are the same as those of the observational result plotted in Fig. \ref{fig:maps12CO32}.}
\label{fig:maps_model_12CO32}
\end{figure*}

\begin{figure*}
\centering
\includegraphics[width=\hsize]{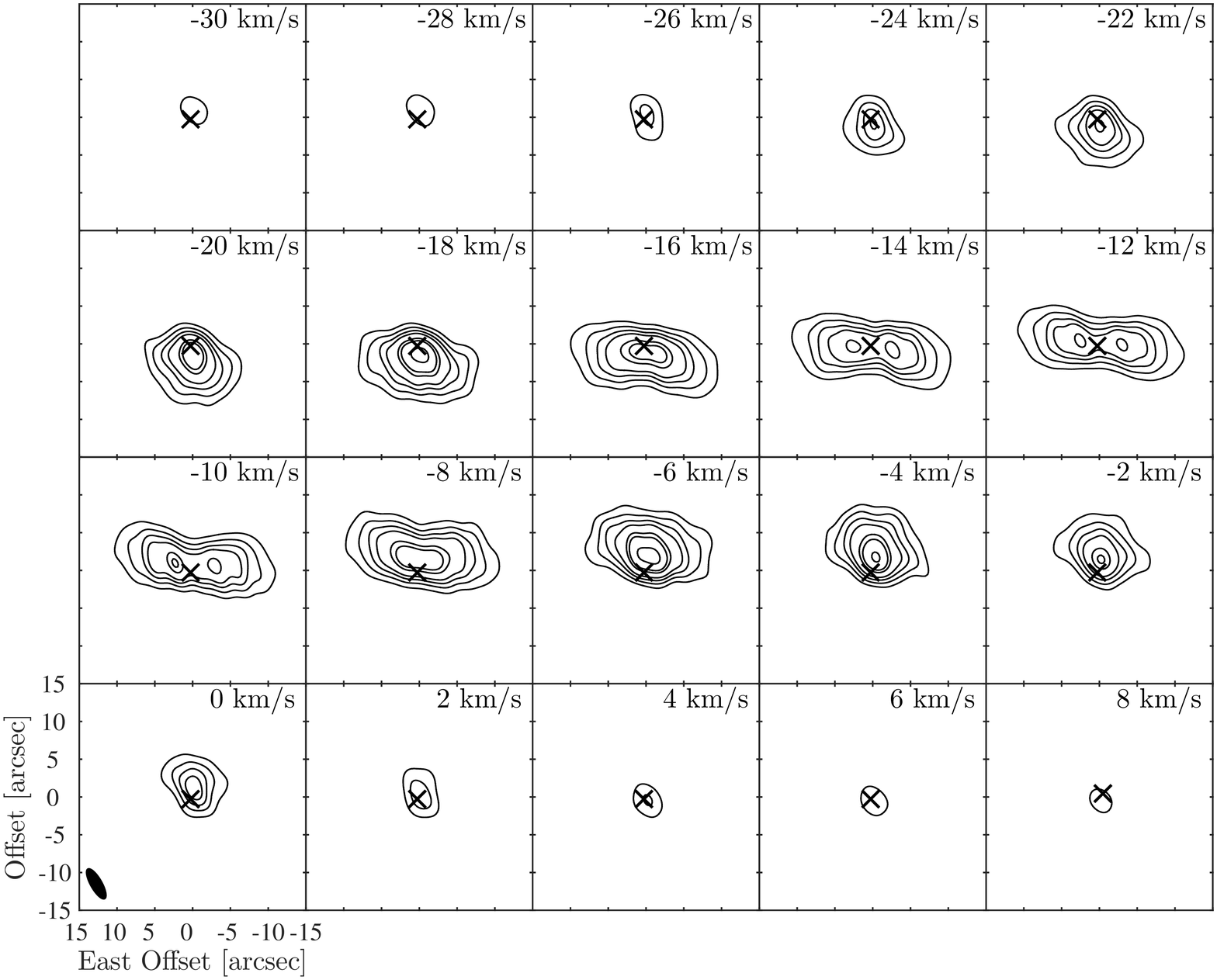}
\caption{Contour maps of the $^{12}$CO $J$=2-1 emission from the model simulating the SMA observation.  The contour levels are the same as those of the observational result plotted in Fig. \ref{fig:maps12CO21}.}
\label{fig:maps_model_12CO21}
\end{figure*}

\subsection{Density and temperature distribution} 
\label{sec:tem&den}
Some simplifying assumptions are necessary to limit the number of free parameters of the 3D model. The density distribution was chosen assuming the following hypothetical, but realistic, scenario for the shaping of the current CSE: The torus (1) is presumably formed in the earlier AGB phase. At some point, the bipolar outflow was triggered and the faster moving material has plowed through the polar regions giving rise to the central component (2), in which the current dynamics are the result of the interaction between the fast outflow and the slower AGB wind. Further out, the bipolar outflow (3) is the faster moving material that has escaped the AGB envelope. The gas density distribution of a spherically expanding AGB envelope created by a constant mass-loss rate, $ \dot{M} $, and expansion velocity,  $ v_{\rm{e}} $, decreases outward according to 
\begin{equation} \label{eq:eqq3}
\centering
n(r)= \frac{\dot{M}}{4\pi  r^{2} v_{\rm{e}} m},
\end{equation}
where m is the particle mass. Since the gas velocity inside the torus depends on the radius according to Eq. (\ref{eq:eqq1}), the H$ _2 $ number density of the torus was set to be proportional to $r^{-3}$, 
\begin{equation} \label{eq:eqq4}
\centering
n_{1}(r,\varphi)=n_{\rm{1a}}\left[ \frac{r}{10^{15} \rm{cm}}\right]^{-3} \frac{1}{f_{\varphi}},
\end{equation}
where $n_{1a} $ is a scaling factor, $ r $ is the radial distance from the center, and the $ f_{\varphi} $ factor is due to the linear dependence of the torus velocity on the latitude (Eq. \ref{eq:eqq1}). The constant velocity of the central component (2) motivates an $r^{-2}$ dependence of the H$ _{2} $ number density,
\begin{equation} \label{eq:eqq5}
\centering
n_{2}(r)=n_{\rm{2a}}\left[ \frac{r}{10^{15} \rm{cm}}\right]^{-2}\rm{.}  
\end{equation}  
The scaling factors, $n_{1a}$ and $n_{2a} $, were initially set using Eq. (\ref{eq:eqq3}) with the mass-loss rate chosen as an average of previous results (1$\times$10$^{-6}$\,M$_{\odot}$\,yr$^{-1}$) and a constant expansion velocity (11\,km\,s$^{-1}$), but then varied to fit the data. 

For bipolar planetary nebulae, a velocity distribution that follows the Hubble law expansion is often found \citep{san2004}. Assuming that the velocity increases radially in the outflow component (3), the density was initially set as an inverse cubic function of the radius,
 \begin{equation} \label{eq:eqq6}
\centering
n_{3}(r)=n_{\rm{3a}}\left[ \frac{r}{10^{15} \rm{cm}}\right]^{-3+\alpha}\rm{,}  
\end{equation}  
and then both the scaling factor $ n_{\rm{a3}}$ and the exponent $ \alpha $ were varied until the data could be reproduced at high velocities (see Sect.~\ref{sec:BestFitModel}). The $^{12}$CO abundance (relative to H$ _{2}$) was assumed to be constant for all three components and a value of 6.5$ \times 10^{-4}$ \citep{knapetal99} was adopted. The $^{13}$CO abundance was changed to fit the $^{13}$CO $J$=3-2 data once all the other parameters had been obtained from fitting the $^{12}$CO lines. 

A description of the gas kinetic temperature as a function of radius in a spherical envelope was presented by \citet{goldetal76}. \cite{chiuetal06} and \cite{knapetal99} successfully applied this temperature distribution when they modeled the torus. A similar dependence on radius, 
 \begin{equation} \label{eq:eqq7}
\centering
T=T_{0} \left[ \frac{r}{10^{15} \rm{cm}}\right] ^{-0.7+\beta} \rm{K}, 
\end{equation} 
was adopted for the whole envelope. The scaling factor, $T_{\rm{0}}$, was varied from 100 K to 500 K, while the exponent was slightly varied around -0.7 through the free parameter $ \beta $.

\subsection{Radiative transfer modeling and imaging} 
\label{sec:rad}
In the radiative transfer model, the parameters introduced in Eqs. (\ref{eq:eqq1})-(\ref{eq:eqq2}) and  (\ref{eq:eqq4})-(\ref{eq:eqq7}), together with the inclination angle of the torus relative to the line of sight, and the PA of the equator were varied until all the available spatially resolved data could be reproduced. The radiative transfer calculation was performed via SHAPE+SHAPEMOL \citep{stefetal11,santetal15}, which is a 3D modeling tool for complex gaseous structures. To solve the radiative transfer equations for $^{12}$CO and $^{13}$CO, the code uses tabulated absorption and emission coefficients that are appropriate for different geometries and kinematic, and calculates the non-LTE level populations via the large velocity gradient (LVG) approximation. We believe the LVG approximation is valid for $\pi^{1}$~Gru because the expansion velocity is larger than the local line width (the thermal contribution is less than 1\,km\,s$^{-1}$) and owing to the large velocity gradient across the CSE. The output simultaneously depends on the gas density, kinetic temperature, velocity distribution, CO isotopologue abundance, and logarithmic velocity gradient ((d$ V $/d$  r$)($ r/V $)). A microturbulent velocity of 2\,km\,s$^{-1}$ \citep{chiuetal06} was adopted. The calculation is conducted for the 17 lowest rotational transitions of the ground-vibrational state of both  $^{12}$CO and $^{13}$CO, including collisions with H$_{2}$ \citep[see][for details]{santetal15}. The envelope was constructed of 128$^3$ grid cells and rendered through a velocity interval of $\pm$100\,km\,s$^{-1}$ around the systemic velocity. The velocity band is divided into 150 channels. 

The comparison between the model and observations requires that the effects of the interferometer, for example, the missing $ u-v $ coverage, noise of the  atmosphere, and electric systems, are included before imaging.  Using the SIMOBSERVE task in CASA, we simulated the visibilities of the observation using the resulting brightness distribution from the radiative transfer calculation. The SMA and ALMA-ACA simulated data has the on-source duration and the antenna configurations of the original observations. The data was then imaged in the same way as the real observational data. The ALMA-TP simulation was performed inside SHAPE+SHAPEMOL by convolving the 3D model with the ALMA primary beam. The ALMA-ACA simulated images were finally combined with the ALMA-TP simulated images using the same procedure as for the real data.

\section{Modeling and interpretation}
\label{sec:ModelResults}
\subsection{Finding the best-fit model}
\label{sec:bestfit}
All free parameters were first adjusted to fit the $^{12}$CO $J$=3-2 line emission. With the many adjustable parameters of the 3D model, the goodness of fit has to be evaluated by taking several different aspects into account. The velocity distribution and inclination angle of the torus were set to reproduce the observed line shapes and spatial distribution seen in the channel maps. The sizes of the different components were constrained by fitting the spatial extent seen in the channel maps (by eye) assuming a distance of 150\,pc \citep{per97}. The density and temperature distribution were primarily chosen to fit of the strength of the emission in the  $^{12}$CO $J$=3-2 line. Those best values were applied to successfully reproduce the $^{12}$CO $J$=2-1. Then the high-$ J $ transition lines $^{12}$CO $J$=5-4 and $J$=9-8 could be fitted by refining the temperature function since the line ratios are sensitive to the kinetic temperature. The $^{13}$CO $J$=3-2 data was matched by only adjusting the $^{13}$CO/H$_{2}$ fractional abundance.  Some additional test models with different velocity fields, morphologies, and temperature distributions (see Appendix A) were considered to find the best-fit model. A chi-square measure was used to evaluate the goodness of the fit from the line profiles, $ \chi ^{2}=\sum_{i=1}^{N}{(I_{i}^{\rm{o}}-I_{i}^{\rm{m}})^{2}/N\sigma_{i}^{2}}$, where $ I_{i}^{\rm{o}} $ and $ I_{i}^{\rm{m}}$ are the beam corrected flux of the observations and the model at channel \textit{i}, respectively, $N$ is the number of channels, and $\sigma $ is the measurement uncertainty of the observed flux, assumed to be 20\% on average.

\begin{table}[th]
\centering
\caption{Input parameters for the best-fit model.}
\label{tab:parameters}
\begin{tabular}{ll}
\hline \hline
 Parameter & Value \\
 \hline 
$R_{\rm{1}}$  & 3$\times$10$^{16}$ cm (2000 AU) \\
$R_{\rm{3}}$  & 5$\times$10$^{16}$ cm (3300 AU)\\
$d$  & 4$\times$10$^{15}$ cm (270 AU)\\
$h$  & 2$\times$10$^{15}$ cm (130 AU)\\
$\varphi_{0}$  & $25^{\circ}  $ \\
$v_{\rm{1a}}$ & 8 km\,s$^{-1}$ \\
$v_{\rm{1b}}$ & 5 km\,s$^{-1}$ \\
$v_{\rm{2}}$ & 14 km\,s$^{-1}$ \\
$v_{\rm{3a}}$ & 11 km\,s$^{-1}$ \\
$v_{\rm{3b}}$ & 89 km\,s$^{-1}$ \\
$ f_{\pm\varphi_{0}}$ & 2.5 \\
$n_{\rm{1a}}$ &  9$\times$10$^{7}$ cm$ ^{-3}$\\
$n_{\rm{2a}}$ &  6$\times$10$^{7}$ cm$ ^{-3}$\\
$n_{\rm{3a}}$ &  6$\times$10$^{7}$ cm$ ^{-3}$\\
$\alpha$ &  0 \\
Inclination\tablefootmark{*} & 40$^{\circ}$ \\
PA of equator & 5$^{\circ}$ \\
$T_{\rm{0}}$ & 190 K \\
$\beta$ & -0.15 \\
$^{12}$CO/$^{13}$CO & 50 \\
 \hline
\end{tabular}
\tablefoot{\tablefoottext{*}{The angle between the line of sight and the equatorial plane}
}
\end{table}
 
\subsection{Best-fit model and comparison to previous results} 
\label{sec:BestFitModel} 

The parameters of the best-fit model are given in Table \ref{tab:parameters}. The outer radius ($R_{\rm{1}}= 3\times$10$^{16}$ cm), and opening angle ($ \varphi_{0} = 25^{\circ}$) of the torus agree with the model by \cite{chiuetal06}. The introduced central component is smaller (angular width $\approx$ 1.7\arcsec) than the synthesized beam of the ALMA-ACA observation and does not affect the fit significantly. At a given radial distance, the gas velocity at the top and bottom edges of the torus is 2.5 times ($ f_{\varphi_{0}}$=2.5) higher than the gas velocity along the equator. An inverse cubic density law (i.e., $\alpha$=0) for the fast bipolar outflow resulted in images very similar to those observed, meaning that the best fit is achieved when the density distribution of the torus and the outflow have the same dependence on radius. Only the scaling factors of the density functions (Eq.~\ref{eq:eqq4} and Eq.~\ref{eq:eqq6}) differ slightly. The best-fit model has a PA of 5$^{\circ}$ and the inclination of the torus relative to the line of sight is 40$^{\circ}$, which agrees with what was found in previous studies \citep{chiuetal06,knapetal99}. If the inclination is decreased (or increased) by more than 10$^{\circ}$ relative to the best-fit value, the model line profiles have a parabolic shape  (or a U-shaped profile with a deep center) in contrast to observed data. If the PA is changed by more than 10$^{\circ}$, the relative intensity of the line peaks is not reproduced. 

The model shows that the temperature distribution suggested by \cite{knapetal99} for the torus cannot be applied to reproduce the high-$  J$ transition lines as well. The temperature in our model is lower ($ T_{0}=190 $ K instead of 300 K) and decreases outward more rapidly ($ \beta = -0.15 $)  than that  found by \cite{knapetal99}.  The model gives a value of 50 for the $^{12}$CO/$^{13}$CO abundance ratio, which is in agreement with upper limit for $\pi^{1}$~Gru derived by \cite{saha92}, but is twice the median value of 25 found for S-type AGB stars \citep{ramsolof14}.

The model line profiles are plotted (Fig.~\ref{fig:Spectrum_comp}) together with the observed line profiles and overall they match the data very well. The SMA $^{12}$CO $J$=2-1 observation only recovered less than 50\% of the total flux as mentioned by \cite{chiuetal06}. Therefore, in the very right panel of Fig.~\ref{fig:Spectrum_comp}, the line intensity has been scaled by a factor of 2 to be able to compare the line shapes. This is only an approximation because the missing flux mainly comes from the extended parts of the envelope with higher velocity, whereas the flux of line core comes from the inner parts and would be better recovered. The peak intensity of the predicted $^{12}$CO $J$=9-8 line is about 20\% less than that from the observation. Our model cannot reproduce the different peak strengths seen in the  $^{12}$CO $J$=9-8 and the $^{13}$CO $J$=3-2 line. This can be an indication that the LVG approximation does not perfectly apply to the torus where the velocity gradient is less steep than in the outflow.

The  spatial distribution of the $^{12}$CO $J$=3-2 and  $J$=2-1 is shown in the model channel maps in Fig. \ref{fig:maps_model_12CO32} and Fig. \ref{fig:maps_model_12CO21}. At the systemic velocity, the model has successfully reproduced the two-peaked distribution seen in the $^{12}$CO $J$=2-1 data without a cavity at the center and the central peak distribution seen in the $^{12}$CO $J$=3-2 data. This means that the model without the central cavity can reproduce the data features. In general, the model images at each channel are very similar to the observations for low to intermediate velocities. The weak bimodal emission distribution at very high-velocity channels (Fig. \ref{fig:maps12CO32}) was not reproduced by our model. As suggested above, the emission could come from the gas at the edges of the two bipolar lobes, but it would depend on the detailed structure and temperature distribution of the outflow, and an exact fit was not attempted.

\begin{figure}
\centering
\includegraphics[width=\hsize]{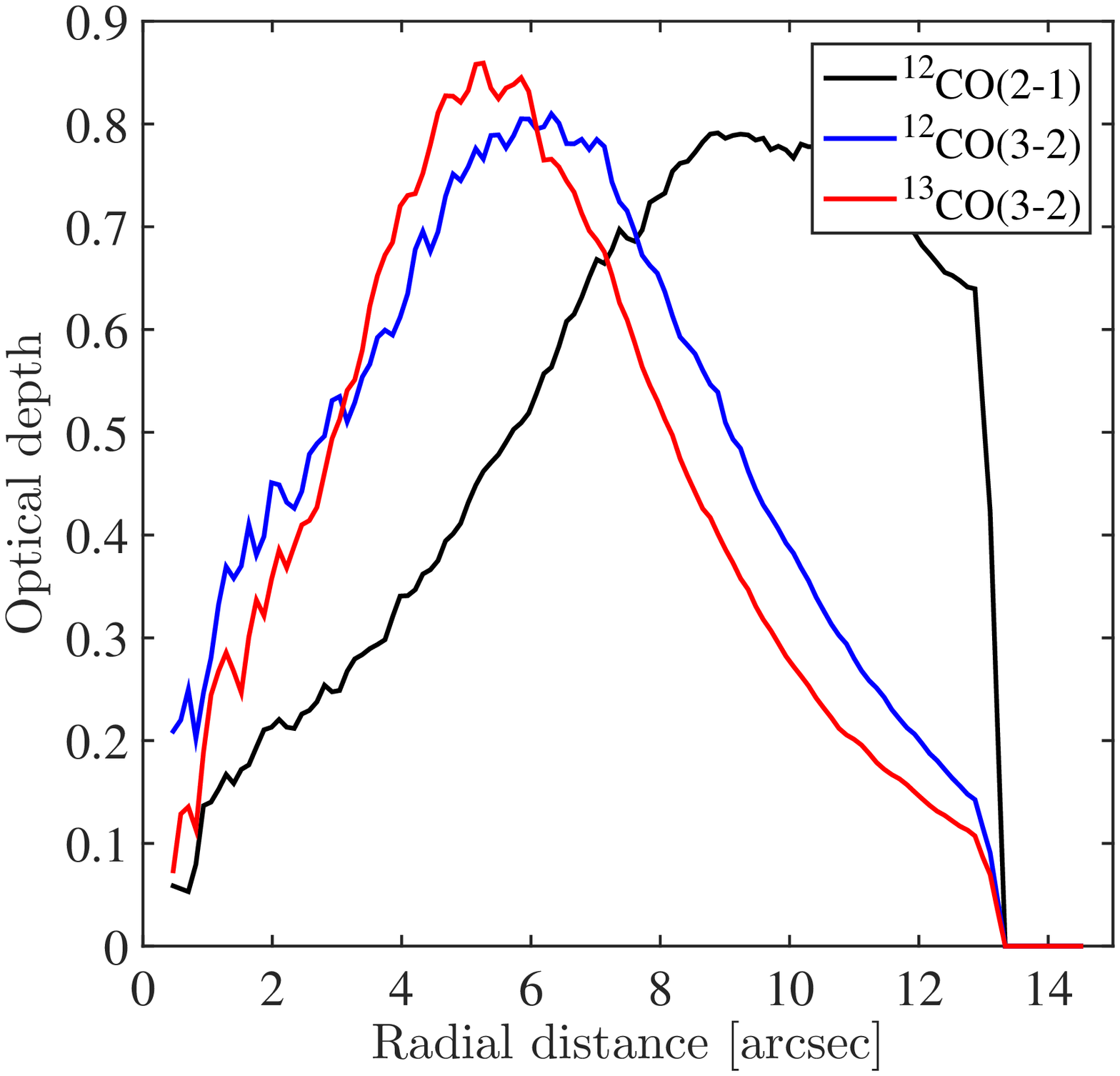}
\caption{Line-of-sight optical depth, calculated from the model, along the equator at the line rest frequencies. The $^{13}$CO $J$=3-2 optical depth was scaled by a factor of 30.}
\label{fig:optical_depth}
\end{figure}

Figure \ref{fig:optical_depth} shows the line-of-sight optical depth (along the system equator at the systemic velocity) as a function of distance from the center for all three modeled lines. Since the emission is optically thin, the peak position of the optical depth indicates where the gas is maximally excited, and agrees with the positions of the emission peaks of the channel map at the systemic velocity (Fig.~\ref{fig:maps12CO21}). The maximum optical depth of the $^{12}$CO $J$=3-2
line occurs inside 5\arcsec and the corresponding two emission peaks are unresolved by the beam.

\subsection{Outflow linear \textit{momentum}} 
\label{sec:momentum} 
The terminal gas velocity along the torus equator is about 13\,km\,s$^{-1}$. If this value is representative of the expansion velocity of the CSE before the formation of the outflow, this corresponds to an average mass-loss rate of about 7.7$\times$10$^{-7}$\,M$_{\odot}$\,yr$^{-1}$. This agrees with the value estimated by \cite{chiuetal06}, and is almost half of that estimated in \cite{knapetal99} when fitting the single-dish $^{12}$CO $J$=2-1 emission. From the sizes of the structures and the expansion velocity distributions, the kinematic timescale of the torus and outflow is about 730 yrs and 160 yrs, respectively. 

The bipolar outflow of the best-fit model suggested has a density distribution that is comparable to the torus at the same radial distance. The estimated total mass of the bipolar outflow is $ M_{\rm{outflow}} $=7.3$\times$10$^{-4}$ M$_{\odot}$ when applying an average particle mass of 3$\times$10$^{-24}$ g and integrating the density distribution across the structure. The highest deprojected gas velocity is about 100 km\,s$^{-1}$ corresponding to a Doppler shift velocity of about 60 km\,s$^{-1}$ in the lines. This results in a linear \textit{momentum} of the outflow, which is the total outflow mass multiplied by the outflow velocity integrated over the velocity distribution, of  $ P_{\rm{outflow}} $=9.6$\times$10$^{37}$ g cm s$^{-1}$. If radiation pressure alone is responsible for lifting the outflow, this process would take about 3300 yrs (calculated from $ P_{\rm{outflow}} $/(\textit{L/c})). This means that the radiation pressure alone would not be sufficient to drive the fast outflow with a kinematic timescale of 160 yrs, which is in agreement with the findings for more evolved sources \citep[e.g.,][]{bujaetal01,olofetal15}.      

\section{Discussion}
\label{sec:dis} 
\subsection{Excitation properties}
All line profiles in this work show a slowly expanding component at the line core and a high-velocity component at the wings. The less extended wings of the $^{12}$CO $J$=5-4 and $J$=9-8 line profiles indicate that the CO molecules are mostly excited to the high-$ J $ states in the central regions where the temperature and density are high enough. The minimum kinetic temperature ($ \sim $250 K) of the $^{12}$CO $J$=9-8 transition, needed for significant collisional excitation, is only reached in regions close to central star. The transition is mainly radiatively excited.

The differences in spatial distribution of the emission from the CO lines are also due to the different excitation requirements. The higher excitation temperature of the $J$=3-2 transition than the $J$=2-1 makes the emission more compact at every channel and particularly at the systemic velocity. Our model can reproduce the different apparent spatial distributions of the two $^{12}$CO lines, and the $^{13}$CO line, reasonably well without including a central cavity \citep[as suggested by][from only having access to the $^{12}$CO $J$=2-1 line]{chiuetal06}. If the cavity actually exists, it must be smaller than the synthesized beam in the case of the $^{12}$CO $J$=3-2 emission. The apparent shape is dependent on the observational setup, line excitation, and actual distribution of the gas, and shows that it is very important to perform detailed radiative transfer modeling before drawing conclusions about the physical distribution of the gas in these types of objects.
 
\subsection{Envelope shaping mechanisms}
The formation of the torus+outflow structure seen in $\pi^{1}$~Gru, poses a challenge to the understanding of stellar evolution theory, as do the typical bipolar structures in PNe. The gravitational perturbation of the known companion of $\pi^{1}$~Gru is not strong enough to concentrate material onto the orbital plane and form the torus \citep[e.g.,][]{mayeetal14}. Even if the orbit is extremely eccentric and the envelope could be compressed at the periastron passage, the timescale for the creation of the torus is too short for it to still be significantly affected. If the orbit is assumed circular, the period would be 6200 yrs to be compared to the kinematic timescale of the torus on the order of 730 yrs. Stellar wind models for AGB stars (assuming a dust-driven wind) typically show wind velocities less than 30\,km\,s$^{-1}$ \citep[e.g.,][]{eriketal14,bladetal15}. For $\pi^{1}$~Gru, the fast ($\leq$100 km\,s$^{-1}$) bipolar outflow has a linear momentum that is higher than the maximum value available from the radiation pressure of the star, which suggests that a different mechanism is required to drive the outflow.

The torus could be created if the star itself is rotating already on the AGB. \cite{dor96} have shown that the effect of a slow rotation (the order of 2\,km\,s$^{-1}$), combined with the strong temperature and density dependence of the dust formation process in AGB stars, can lead to an enhanced equatorial mass loss and produce an elliptical envelope. By applying the wind compressed disk model to AGB stars,  \cite{Ign96} showed that the coriolis effect, effective if the star is rotating fast enough, can produce a disk-like structure. A close companion or a binary merger can spin-up AGB stars to the rotation rate required for the wind-compressed disk to be formed. However, there is no evidence for rotation observed in $\pi^{1}$~Gru, but a velocity below 2\,km\,s$^{-1}$ cannot be excluded. The gravitational effect of a close companion star, or a giant planet, can play an important role in the formation of a torus \citep{iben93}.

Whether single stars can form bipolar morphologies was initially investigated in interacting wind models \citep[see, e.g.,][and references therein]{balifran02}. In these models, a fast isotropic wind is launched inside a slowly expanding torus (without investigating the formation of the torus itself) with a density contrast between the pole and the equator. A hot bubble is created in the post-shock gas and the bubble expands at constant pressure. The expansion velocity of the bubble depends on the density distribution of the torus and varies inversely from the pole to the equator \cite[e.g.,][]{icke88}. The very high gas velocity at the poles eventually launches the bipolar outflow. Wind-interaction models including detail hydrodynamics and microphysics \citep[e.g.,][]{frank94} have confirmed the dependence of the shaping on the density distribution of the previously ejected gas envelope. A density contrast (between the torus equator and the pole) from 2 to 5 results in a fast bipolar outflow. A higher value results in a highly collimated outflow. Within the modeled torus, the best-fit model of $\pi^{1}$ Gru has a density contrast of 2.5 between the equator and the edge at an angle of 25$^{\circ}$. If a linear dependence with angle is assumed, the contrast between the equator and the pole is 6.4, which is a reasonable agreement between the results from the interacting wind models and the observed morphology.

The kinematic timescales for the torus and fast outflow in our model are similar to the typical values found for PPNe and PNe \citep{bujaetal01}, showing the same circumstellar structures. The momentum excess (compared to what is available from radiation pressure alone) implies the need for an additional wind driver. Wind formation and the collimation of bipolar outflows have also been studied using magneto-hydrodynamics (MHD) models where the wind is driven by magnetic pressure \citep{garc99,garc05}. These models are successful in creating bipolar structures and highly collimated jets when the magnetic field is strong enough and the rotational velocity is sufficient. However, recent investigations combining the results of MHD and stellar evolution models, have shown that the rotational velocities retained at the end of the AGB are not sufficient to form bipolar PNe \citep{garc16}. \cite{matt00} used MHD models to show that a modest magnetic field \citep[of order a few G at the surface, similar to what has been measured on some AGB stars;][]{lebre14} is sufficient to form a dense equatorial disk around an AGB star with a slow, massive wind. In combination with the interacting wind models described above, this gives another possible explanation for the observed morphology. Again, the origin of the magnetic field or the launching of the fast wind are not explained, but would require an extra source of angular momentum, e.g., a binary companion. 

The suggested second companion remains to drive the formation of the observed morphology. If close enough, it could also accrete the wind through an accretion disk and possibly drive the fast outflow. The low velocity of the torus sets favorable conditions for wind Roche-lobe overflow \citep[wRLOF;][]{moh12}, where a slowly expanding wind first fills the Roche lobe of the primary star (the AGB star) and then flows through the inner Lagrangian point onto the accretion disk of the companion. Models including the wRLOF scenario can produce substantial accretion rates also in larger separation binaries \citep[e.g., Mira, R Scl,][]{ramsetal14,maeretal12}, where traditional Bondi-Hoyle accretion models fail to reproduce the observations. The exact requirements to drive an outflow with the momentum observed in $\pi^{1}$ Gru will be the subject of a future publication.
   
The CSE of $\pi^{1}$~Gru is not the only case where a torus plus bipolar structure has been observed around an AGB star. The CO line observations of V Hya also shows a similar structure \citep{sahai03}. Using the CO $J$=2-1 and 3-2 line observation,  \cite{hir04} have distinguished three components in the CSE of V Hya based on its kinematic properties, which are similar to the results in this study. These authors also suggested a similar explanation for generating an intermediate-velocity component as the central component in our model. Another similar example is the post-AGB star, CRL 618, with a very fast, collimated outflow rising from a low-velocity, dense core \citep{san04}. Owing to the complex geometry showing features commonly found in stars of the next evolutionary phase, $\pi^{1}$~Gru is an extremely interesting case to study in order to find the missing link between the spherical outflows of AGB stars and the bipolar outflows observed at later stages. 

\section{Summary}
\label{sec:sum}
We have presented the analysis of new ALMA-ACA data of the $^{12}$CO $J$=3-2 and $^{13}$CO $J$=3-2 line emission, together with previously observed $^{12}$CO $J$=2-1 data from the S-type AGB star $\pi^{1}$~Gru. The high-sensitivity ALMA observations recovered the extended emission, and for the first time, resolved the high-velocity component. The analysed data, including low-$ J $ transitions (from ALMA and SMA observations) high-$ J $ transition (from Herschel/HIFI observations) provided sufficient constraints for a 3D radiative transfer model. The best-fit model reconstructing the gas envelope has satisfactorily reproduced the line profiles, channel maps, and suggested a reasonable value for the abundance of $^{12}$CO/$^{13}$CO. The gas envelope is modeled as a system of three separate components: a radially expanding torus with the velocity linearly increasing with latitude and radial distance, a central,  radially expanding component that may have resulted from the dynamical interaction between the fast outflow and the torus, and a fast bipolar flow perpendicular to the equator with a radially expanding velocity field. The outflow momentum excess found in our model rules out a scenario in which radiation pressure alone can lift the high-velocity outflow. The density contrast between the equator and polar regions suggested from various formation mechanisms can successfully reproduce the data. This supports that the gravitational effect of a close companion is involved the torus formation, while the wind interaction mechanism and/or a bipolar magnetic field could be included when considering the outflow formation.

\begin{acknowledgements}
The authors would like to thank the staff at the Nordic ALMA ARC node for their indispensable help and support. This paper makes use of the following ALMA data: ADS/JAO.ALMA\#2012.1.00524.S. ALMA is a partnership of ESO (representing its member states), NSF (USA) and NINS (Japan), together with NRC (Canada), NSC and ASIAA (Taiwan), and KASI (Republic of Korea), in cooperation with the Republic of Chile. The Joint ALMA Observatory is operated by ESO, AUI/NRAO, and NAOJ.\\
We are grateful to T. Danilovich  for helping us with the Herschel/HIFI data.\\
S. Mohamed is grateful to the South African National Research Foundation (NRF) for a research grant.\\ 
W.~H.~T. Vlemmings acknowledges support from ERC consolidator grant 614264.\\ 
C. Paladini is supported by the Belgian Fund for Scientific Re- search F.R.S.- FNRS. 
\end{acknowledgements}

\bibliographystyle{aa}

\bibliography{pi1Gru}

\begin{appendix}

\section{Finding the best-fit model}
\label{app:find}
Some alternative models that were tested to find the best-fit model are presented here. The models with different velocity fields, morphologies, and temperature distributions were constrained by both the channel maps and line profiles.

A model using a constant expansion velocity for the whole torus could not reproduce the line profile nor the channel maps. Indeed the line profile of the $^{12}$CO $J$=3-2 line from such a model has an intensity at the systemic velocity that is much smaller than that of the peaks (at about $ \pm $6 km\,s$^{-1}$), while the center and peak intensities are comparable in the observed line profile. The model channel maps are rather sensitive to the assumed velocity field. If $v_{\rm{1}} $ is reduced by 10\%, the $^{12}$CO $J$=2-1 emission in the velocity channel beyond $\pm$12\,km\,s$^{-1}$ is reduced below the noise level. The best-fit velocity distribution is also dependent on the chosen inclination angle of the torus, which was constrained by the flattened shape of the torus seen at the systemic velocity in the channel maps (Fig. \ref{fig:maps12CO32}). Finally, the PA of the torus was set to reproduce the relative intensity of the two peaks compared to each other in the observed line profiles (see Section \ref{sec:BestFitModel}).  

A collimated velocity field, which is typically found in P-PNe and PNe in which the outflow gas velocity is perpendicular to the equatorial plane, was first attempted to model the fast bipolar component.  This gave line profiles similar to the observed profiles, however, it also resulted in a very large intensity ratio between the central velocity channel and the nearby channels, which is not seen in the observed images.

An alternative model, without the central component (2), was also tested. In that model, the torus reaches regions close to the central star and is shaped like a flared disk. The fast bipolar outflow rises on top of the torus. This model successfully reproduced the spatial distribution of the $^{12}$CO $J$=2-1 and the $^{12}$CO $J$=3-2 emission, in particular the two-peaked distribution of the $^{12}$CO $J$=2-1 emission, as well as the single central peak of the $^{12}$CO $J$=3-2 emission at the systemic velocity (without a central cavity). However, in order to reproduce the data, the innermost outflow velocity had to be lower than the velocity of the torus at the same position, which would be unphysical. 

The temperature distribution, as suggested by \cite{knapetal99}, results in a good agreement with the observations in the low-$  J$ transition lines, the $^{12}$CO $J$=2-1 and $J$=3-2, and $^{13}$CO $J$=3-2 line. However, it overestimates the intensity of the $^{12}$CO $J$=5-4 and $J$=9-8 lines that are excited in the central regions. The two lines are sensitive to temperature and can be simultaneously fitted when the temperature decreases outward more rapidly than that found in  \cite{knapetal99} (see Section \ref{sec:tem&den}).

\end{appendix}
\end{document}